\documentclass{Interspeech2024}

\usepackage{multirow}
\usepackage{booktabs}
\usepackage{algorithm}
\usepackage{algorithmicx}
\usepackage{algpseudocode}

\interspeechcameraready

\title{DINO-VITS: Data-Efficient Zero-Shot TTS with Self-Supervised Speaker Verification Loss for Noise Robustness}

\small \name[affiliation={1}]{Vikentii}{Pankov}
\name[affiliation={1}]{Valeria}{Pronina}
\name[affiliation={2}]{Alexander}{Kuzmin}
\name[affiliation={2}]{Maksim}{Borisov}
\name[affiliation={3}]{Nikita}{Usoltsev}
\name[affiliation={5}]{Xingshan}{Zeng}
\name[affiliation={1}]{Alexander}{Golubkov}
\name[affiliation={1}]{Nikolai}{Ermolenko}
\name[affiliation={4}]{Aleksandra}{Shirshova}
\name[affiliation={1}]{Yulia}{Matveeva}

\address{
\small
  $^1$Huawei Technologies, Russia
  $^2$ITMO University, Russia
  $^3$HSE, Russia
  $^4$SPbU, Russia
  $^5$Huawei Noah Ark Lab, China}
\email{vkpankov@gmail.com, yulia.matveeva@yahoo.com}


\begin{document}

\maketitle

\begin{abstract}
    
We address zero-shot TTS systems' noise-robustness problem by proposing a dual-objective training for the speaker encoder using self-supervised DINO loss. This approach enhances the speaker encoder with the speech synthesis objective, capturing a wider range of speech characteristics beneficial for voice cloning. At the same time, the DINO objective improves speaker representation learning, ensuring robustness to noise and speaker discriminability. Experiments demonstrate significant improvements in subjective metrics under both clean and noisy conditions, outperforming traditional speaker-encoder-based TTS systems. Additionally, we explore training zero-shot TTS on noisy, unlabeled data. Our two-stage training strategy, leveraging self-supervised speech models to distinguish between noisy and clean speech, shows notable advances in similarity and naturalness, especially with noisy training datasets, compared to the ASR-transcription-based approach.

\end{abstract}

\section{Introduction}

In this paper, we focus on zero-shot TTS, meaning that the system should be capable of cloning an unseen reference voice without additional training, and consider two types of challenging settings for such systems. Firstly, we focus on the problem of the robustness of zero-shot TTS systems against background noise present in target speaker reference audios at the inference stage, a common scenario in a real-world environment. Secondly, we investigate the problem of training these systems using noisy, unlabeled data, which represents a considerable part of real-world audio data.

{\bf Robustness to noisy reference audios at inference.} In \cite{norespeech_2022_2023}, a  probabilistic denoising diffusion model is trained to extract noise-agnostic style embeddings from reference audios and a wav2vec2.0-based speaker encoder is trained to generate speaker embeddings. Noise augmentations are used to train a zero-shot TTS system such that the style features extracted from noisy mel-spectrograms are close to those extracted from non-noisy ones. Some works, such as \cite{cdat_2020}, utilize adversarial training to ensure control over noise being encoded or not encoded in the intermediate representations of the models. In \cite{cross_lingual_expressive_ai_dubbing}, the authors train a speech synthesis model in which the robustness to noises in references is achieved via two independent methods. One is denoising the reference audios both at inference and at training with an external denoiser model. The second method consists in introducing an additional noise encoder disentangling the noise information in reference audio from useful prosodic information and feeding a fixed clean audio to the noise encoder at inference. In \cite{byol_a_vc}, the authors propose to use pretrained self-supervised BYOL-A features for speaker conditioning, and focus on augmentation strategies for this model pretraining. In \cite{fujita2024noiserobust}, a self-supervised WavLM model is utilized as a speaker encoder, focusing on a parameter-efficient finetuning methods to improve the noise-robust properties of the WavLM embedding extractor through noise augmentation of input.

In contrast to the mentioned works, we propose a joint training strategy for the speaker encoder part of our TTS system using self-supervised DINO loss \cite{DINO_2023} and reconstruction loss. Compared to \cite{norespeech_2022_2023,cdat_2020,cross_lingual_expressive_ai_dubbing}, our method does not use external denoising models or additional noise encoders. Compared to \cite{fujita2024noiserobust}, we utilize a compact jointly-trained speaker encoder model, leverage large datasets during training, and evaluate model performance on real-life noisy data.

{\bf Training from noisy untranscribed data.} 

In the existing literature, unsupervised TTS approaches fall into one of two categories: transfer learning from self-supervised audio-representation models \cite{transfer_from_wav2vec_unlabeled_2022,unsuptts_2022} and offline voice conversion systems trained on untranscribed data to re-synthesize a transcribed corpus with additional voices \cite{vc_for_vc_2022}. The work \cite{transfer_from_wav2vec_unlabeled_2022} utilizes features from wav2vec 2.0, which are converted into discrete units via clustering, as content inputs for training a VITS model \cite{vits2021}. Further, it replaces the unit encoder with a phoneme encoder during the phoneme-based finetuning phase. In \cite{unsuptts_2022}, an unsupervised ASR system is proposed that converts wav2vec 2.0 features into phoneme sequences for generating pseudo-transcripts for an untranscribed dataset.

Approaches to tackle noise include conditioning on noise information to enable synthesis from noisy data with minimal quality degradation and applying pre-trained denoisers to noisy recordings \cite{cross_lingual_expressive_ai_dubbing}. In Rep2wav \cite{zhu2023rep2wav}, the authors separately train FastSpeech-based model for mapping text to WavLM features, using a speech enhancement model, and separately train single-speaker vocoder to synthesize speech from WavLM output. 

Our voice cloning system leverages the self-supervised audio representations' capacity, as shown by the HuBERT model~\cite{dehubert_2023}, to intrinsically differentiate between noisy and clean speech. During training, the model learns to associate noisy audio inputs with corresponding noisy references and targets, and clean inputs with clean targets.

\textbf{We summarise our contributions as follows:}
\begin{itemize}

    \item \textbf{Robustness to noisy reference audios at inference.} We develop a multi-task training method for the speaker encoder in our voice cloning framework, leveraging the self-supervised DINO loss \cite{DINO_2023}. This strategy allows the speaker encoder to better capture a wider range of speech characteristics in the generated embeddings through the voice cloning objective.  At the same time, speaker representation learning is enhanced through the DINO objective, which ensures that the system maintains speaker discriminability and robustness across diverse vocal conditions, including background noise. Our experiments show that this approach can bring substantial improvements in naturalness and speaker similarity in both clean and especially real-life noisy scenarios, outperforming traditional AAM-Softmax-based training methods.

\item \textbf{Training on noisy untranscribed data.} We investigated the inherent capabilities of self-supervised learning HuBERT model~\cite{dehubert_2023} to differentiate between noisy and clean speech without prior noise-specific training. Unlike traditional methods requiring explicit noise labeling or the incorporation of noise-aware mechanisms, our system inherently conditions on the noise in training data. We utilize this property for semi-supervised noise-robust training of our voice cloning model and show that it significantly outperforms ASR transcription-based approach.
    
\end{itemize}

\section{Method}
\label{sec:method}
\subsection{System architecture}
\label{ssec:method_vc2}

\begin{figure}
  \centering
  \includegraphics[width=0.9\linewidth]{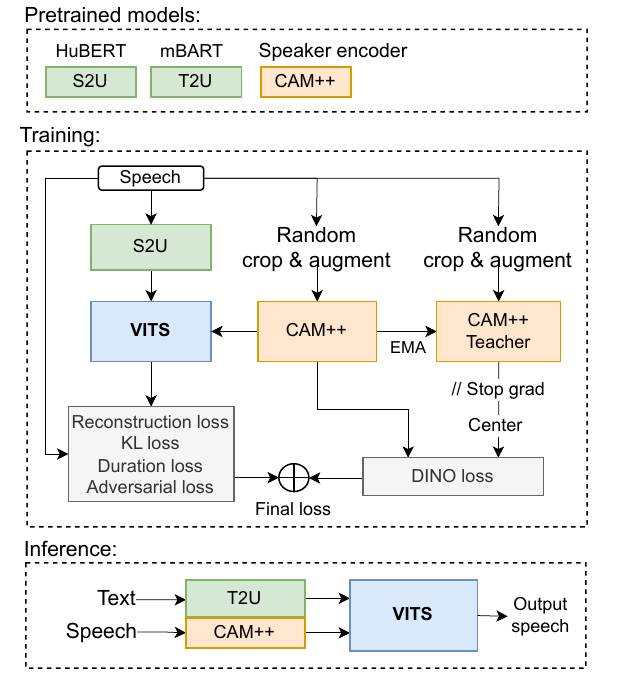}
  \caption{Architecture of the proposed method (DINO-VITS). The CAM++ Teacher is an exponential moving average (EMA) of CAM++ weights. The Center operation subtracts EMA of previous speaker embeddings from teacher output.}
  \label{mainscheme}
\end{figure}

Our system, as illustrated in Figure 1, is comprised of four modules: 1) a pretrained HuBERT-based S2U (Speech-to-Unit) module that converts content speech into a sequence of discrete symbolic representations, 2) an mBART-based T2U (Text-to-Unit) module which maps content text to the same representation space, learning the sequence-to-sequence task of predicting HuBERT units from ARPABET phonemes, 3) a speaker verification CAM++ model, also pretrained, used as a speaker encoder, and 4) a VITS architecture (U2S - Unit-to-Speech) that accepts units as linguistic content input from either the S2U or T2U module along with a speaker embedding, to synthesize the given content in a voice similar to the one presented in the reference audio fed into the speaker encoder. During training, we utilize features from the S2U module. During inference, it is replaced by the T2U module, which is separately trained on a clean subset of training data to predict the outputs of the pretrained HuBERT S2U. Our U2S module is a VITS model with conditioning on the speaker embedding of the posterior encoder, the flow, and the duration predictor. The number of parameters is 95 million in HuBERT, 610 million in mBART, 40 million in VITS, and 6 million in the speaker encoder. The total inference RTF is 0.45 on NVIDIA RTX 2080 GPU.

\subsection{Jointly trained speaker encoder and robustness to noisy reference audios at inference}
\label{ssec:method_speaker_verification}

We utilize a pretrained speaker verification CAM++ model \cite{campplus2023} denoted by $SE(\cdot)$ for zero-shot voice cloning, leveraging the speaker-rich Voxceleb2 dataset during the pretraining stage. As with most state-of-the-art speaker verification models, it is trained with the AAM-Softmax loss, which encourages compact clustering of embeddings for the same speaker while pushing apart embeddings from different speakers. Consequently, the speaker embeddings from the pretrained model are not fully suitable for the speech synthesis task due to their limited style adaptability: speaker verification generally benefits from emotion and speech style invariance, whereas voice cloning requires it for enhancing style transfer capability.

To address the limitations of the pretrained speaker encoder, we fine-tune it jointly with the speech synthesis model during training. Simply unfreezing the speaker encoder model during TTS training leads to catastrophic forgetting of its initial useful properties, such as high speaker discriminability and robustness to noise. To circumvent these issues, we employ noise augmentations of reference audios and introduce multi-task learning of the speaker encoder with the self-supervised DINO loss \cite{DINO_2023}. Compared to AAM-Softmax, DINO affords a more flexible embedding space that captures within-speaker variations such as style and emotion. This is achieved by minimizing the cross entropy loss $\mathcal{L}_{\text{DINO}}$ (\ref{dinoeq}) between the output distributions of the teacher $P_T$ and student $P_S$ networks over different augmented random crops $x_{a1}$ and $x_{a2}$ of the same speech input $x$, without explicitly enforcing tight clustering within speakers, in contrast to the supervised loss AAM-Softmax, as shown below.

\begin{equation}
    \mathcal{L}_{\text{DINO}} = -\sum_{i=1}^{K} \sigma\left(\frac{P_T(x_{a1})_i - C}{\tau}\right) \log \sigma\left(\frac{P_S(x_{a2})_i}{\tau}\right),
    \label{dinoeq}
\end{equation}
where $\sigma$ is the softmax function, $C$ is a center vector computed as an exponential moving average of previous teacher outputs, and $\tau$ is a temperature parameter. The student network $P_S$ is represented by the speaker encoder $SE$ coupled with a three-layer projection head that produces a $K$-dimensional output. The weights of $P_T$ are maintained as an exponential moving average of the $P_S$ weights.

We summarize the training and inference steps of our voice cloning system in Algorithm \ref{mainalg}.

\begin{algorithm}
\caption{Proposed Speech Synthesis System}
\begin{algorithmic}[1]
\State \textbf{Stage 1: Pretraining}
\State Pretrain Speaker Encoder $SE(\cdot)$ on a multi-speaker dataset.
\State Pretrain HuBERT $H(\cdot)$ to map speech $x$ to a hidden unsupervised representation $h$.
\State Pretrain mBART $M(\cdot)$ to map phonemes to HuBERT units.

\State \textbf{Stage 2: Training}
\State Discretize HuBERT output $h$ using k-means into 1000 clusters and remove consecutive duplicates in all training records to obtain a sequence $u$ for each training audio.
\State Extract two augmented random crops $x_{a1}$ and $x_{a2}$ from $x$ 
\State Extract speaker embedding $e = SE(x_{a1})$ and  compute $\mathcal{L}_{\text{DINO}}$ as described in Section 3.1 of paper \cite{DINO_2023} 
\State Compute $\mathcal{L}_{vae}$ as described in Section 2.4 of  paper \cite{vits2021} 
\State Perform training step, with $u$ as a prior encoder input and with $e$ for speaker conditioning, to jointly optimize speech synthesis model weights $\phi$ and speaker encoder weights $\theta$:
$$ 
\min_{\theta,\phi} \left( \mathcal{L}_{vae}(\theta, \phi; u,e,x) + \lambda \mathcal{L}_{\text{DINO}}(\theta;x_{a1},x_{a2}) \right)
$$

\State \textbf{Stage 3: Inference}
\State Obtain the unit sequence $u = M(\text{text})$ using the pretrained mBART, and obtain the speaker embedding $e$ using the jointly trained speaker encoder.
\State Synthesize speech using VITS by inputting $u$ and $e$.
\end{algorithmic}
\label{mainalg}
\end{algorithm}

\section{Experiments and results}
\label{sec:experiments}

\subsection{Experiment setup}
\label{sec:gen_decsription}

\subsubsection{Data}
We utilize VCTK \cite{vctk_2019}, LibriTTS \cite{libritts_2019}, and LibriLight \cite{librilight_2020} datasets for training. We randomly selected records less than 8 sec, maximum of 3000 records per speaker. We apply loudness normalization and downsampling to 16kHz for all datasets.

We augment the reference input with noises randomly chosen from the MUSAN dataset while the target audio always remains clean. This augmentation is applied with a 50\% probability and with random SNR (16-25 for babble noise composed of 1-6 random speech utterances, 6-20 for music, and 3-20 for noise) to each training record in each batch.

The evaluation was conducted on a ChiME3 \cite{chime3} subset, featuring 8 speakers and 15 reference audios per speaker, roughly evenly distributed across four noisy recording locations. For each reference, a different ground truth (GT) source audio and text were selected from the same speaker. Clean reference audios and GT recordings were chosen from the clean ``booth" environment. All test reference audios were trimmed to 3 seconds. Subjective assessments of quality (naturalness) and speaker similarity were performed on the Toloka crowd-sourcing platform~\cite{toloka}, with 10 annotators rating each audio.

\subsubsection{Hyperparameters}
The speaker encoder  was pretrained on VoxCeleb2 with noise augmentations from MUSAN and RIRS \cite{rirs} datasets with all hyperparameters replicated after the  original work \cite{campplus2023}. For S2U we take a pretrained HuBERT model with k-means clustering \cite{mhubert_base} to generate discretized features (units). For training the T2U module we utilize the unit-based {mBART} checkpoint \cite{mbart_large} and fine-tune it on the English part of LibriSpeech \cite{librispeech_2015}. 

The architecture and training hyperparameters of VITS module are equal to original VITS \cite{vits2021}, except the added speaker encoder, DINO loss, augmentations for reference input, and increased reference embedding size from 192 to 256. All models were trained on two NVIDIA RTX 3090 GPUs with total batch size 80 for 5 days. We trained them in two stages: a 95k-iteration pretraining stage with a frozen speaker encoder (except the last layer), followed by 175k iterations with an unfrozen speaker encoder. In contrast to the DINO for speaker verification in \cite{DINO_2023}, which employed multiple local and global segments, we used only two random speech segments $x_{a1}$ and $x_{a2}$ for each utterance. We set the segment size as $min(s,5)$ sec, where $s$ is the minimal speech duration in the current batch. We also use MUSAN augmentations only. The other hyperparameters of DINO are replicated after  \cite{DINO_2023}.

\subsection{Robustness to noises at inference}

\subsubsection{Style preservation in speaker embeddings}

To verify the hypothesis regarding our approach's enhanced capability to encode style in reference embeddings, we developed an emotion recognition classifier on speaker encoder outputs. This classifier comprises two fully connected linear layers, with the initial layer configured as $256 \times 128$ for the jointly trained speaker encoder and $512 \times 64$ for the pretrained CAM++ model. We estimated emotion recognition accuracy using 5-fold cross validation on two datasets: CREMA-D \cite{cao2014crema} and IEMOCAP \cite{busso2008iemocap}. For CREMA-D, jointly trained model yielded 62.4\% accuracy (±2.2), while pretrained CAM++ yielded 53.4\% (±1.0). For IEMOCAP, our model reached 45.8\% accuracy (±1.7), and pretrained CAM++ yielded 39.8\% (±2.3). A notable increase of up to +9\% in accuracy after the multi-task training with DINO loss indicates that proposed joint training framework more effectively preserves the encoding of style information in the reference embedding when compared to the original pre-training with AAM-Softmax loss for speaker verification.

\subsubsection{Baselines}

We prepare a unit-based TTS baseline inspired by \cite{byol_a_vc}. While the original uses a non-attentive Tacotron and LPCNet, we incorporate the BYOL-A encoder into our end-to-end VITS architecture. The encoder remains frozen during TTS training, as described in \cite{byol_a_vc}.

For the YourTTS baseline we reproduce the model and training hyper-parameters from the original paper \cite{yourtts_2022} without multi-lingual synthesis part.
Since originally YourTTS was not specifically intended to work in noisy conditions, we add a DEMUCS \cite{defossez2021hybrid} denoiser to provide a stronger baseline ({\bf YTd} in Table \ref{tab:baselines}) for tests with noisy reference audios.

\begin{table}[h]
\caption{Comparison of the proposed DINO-VITS method ({Ours}) to YourTTS ({\bf YT}), DEMUCS denoiser+YourTTS ({\bf YTd}), and BYOL-A ({\bf BY}) for clean and noisy test subsets}
\label{tab:baselines}
\centering
\setlength{\tabcolsep}{4pt} 
\renewcommand{\arraystretch}{1.0} 
\resizebox{\columnwidth}{!}{
\begin{tabular}{llllll}
\toprule
 & \multicolumn{2}{c}{Naturalness} & \multicolumn{2}{c}{Similarity} \\
 & Clean & Noisy & Clean & Noisy \\
\midrule
GT & 4.68 $\pm$ 0.03 & - & 3.94 $\pm$ 0.07 & - \\
Ours & \textbf{4.00 $\pm$ 0.05} & \textbf{3.55 $\pm$ 0.10} & \textbf{3.85 $\pm$ 0.08} & \textbf{3.52 $\pm$ 0.08} \\
YT & 3.96 $\pm$ 0.05 & 3.11 $\pm$ 0.11 & 3.33 $\pm$ 0.08 & 3.20 $\pm$ 0.08 \\
YTd & - & 3.28 $\pm$ 0.10 & - & 3.35 $\pm$ 0.08 \\
BY & - & 1.85 $\pm$ 0.09 & - & 1.89 $\pm$ 0.07 \\
\bottomrule
\end{tabular}
}
\end{table}

As shown in Table \ref{tab:baselines}, DINO-VITS significantly outperforms both methods in clean (by similarity) and noisy conditions (by similarity and naturalness). These results demonstrate the advantage of our proposed multi-task training approach over using a pre-trained speaker encoder in supervised mode (YourTTS) or self-supervised mode (BYOL-A).

\subsubsection{Ablation studies} We conduct several ablation studies on the role of the choice of speaker verification loss and data augmentation. The results are reported in Table \ref{tab:ablations}. We fixed all hyperparameters except for the speaker verification loss and the reference audio augmentations: DINO-VITS ({\bf Ours}) is the proposed method with MUSAN noises added to reference audios at random during training and with DINO speaker verification loss, in {\bf AV}, DINO loss is replaced by AAM-Softmax loss, and in {\bf NV}, we further remove noise augmentations of references (keeping multi-task learning with AAM-Softmax loss).

\begin{table}[h]
\caption{Ablations for the proposed method ({\bf Ours}) tested with clean references. {\bf AV}: AAM-Softmax instead of DINO loss. {\bf NV}: AV + remove augmentations}
\label{tab:ablations}
\centering
\setlength{\tabcolsep}{4pt} 
\renewcommand{\arraystretch}{1.0} 
\resizebox{\columnwidth}{!}{
\begin{tabular}{ll|ll|ll}
\toprule
 & \multicolumn{2}{c|}{Naturalness} & \multicolumn{2}{c}{Similarity} \\
 & Clean & Noisy & Clean & Noisy \\
\midrule
GT & 4.58 $\pm$ 0.04 & - & 3.83 $\pm$ 0.08 & - \\
Ours & 4.03 $\pm$ 0.04 & \textbf{4.07 $\pm$ 0.05} & \textbf{3.88 $\pm$ 0.06} & \textbf{3.50 $\pm$ 0.07} \\
AV & 4.00 $\pm$ 0.04 & 3.58 $\pm$ 0.05 & 3.73 $\pm$ 0.07 & 3.23 $\pm$ 0.07 \\
NV & \textbf{4.04 $\pm$ 0.04} & 2.47 $\pm$ 0.05 & 3.86 $\pm$ 0.06 & 2.50 $\pm$ 0.08 \\
\bottomrule
\end{tabular}
}
\end{table}

The results of the ablation study in Table \ref{tab:ablations} indicate that  DINO loss performs similarly to AAM-Softmax in clean conditions. However, DINO significantly enhances both naturalness and similarity for noisy scenario when compared to AAM-Softmax. This confirms the ability of DINO loss to preserve noise-robust properties of speaker encoder during its joint training with voice cloning model. 

\subsection{Training from noisy untranscribed data}
\label{ssec:exp_corrupted_units}


We test the robustness of HuBERT-based approach to be trained on unlabeled and noisy data and compare it with the traditional ASR-based method. To do so, we select Whisper medium model~\cite{whisper_2022} – a large pretrained speech recognition model, that maps untranscribed audio to phonemes, -- and use these phonemes as inputs for U2S. Both models are trained with a noisy variant of the LibriLight subset, obtained by adding noises from the MUSAN dataset~\cite{musan2015} to a randomly chosen 40\% of the content records with a resulting SNR of 0 dB for each augmented record, and original variants of LibriTTS and VCTK.


\subsubsection{Results}


\begin{table}[htbp]
\caption{Comparison on clean target audios for models trained on clean (-C) and noisy (-N) training data.}
\label{tab:clean_data}
\centering
\begin{tabular}{lccc}
\toprule
Model & Naturalness & Similarity & CER \\
\midrule
GT & 4.72$\pm$0.03 & 4.24$\pm$0.07 & 3.86$\pm$0.20\\
Whisper-C & 3.75$\pm$0.05 & 3.43$\pm$0.08 & 6.84$\pm$0.35 \\
Ours-C & \textbf{4.01}$\pm$\textbf{0.04} & \textbf{3.48}$\pm$\textbf{0.08} & \textbf{4.74$\pm$0.26} \\
Whisper-N & 3.60$\pm$0.05 & 3.26$\pm$0.08 & 9.99$\pm$0.53 \\
Ours-N & \textbf{4.04}$\pm$\textbf{0.04} & \textbf{3.50}$\pm$\textbf{0.08} & \textbf{4.69$\pm$0.25} \\
\bottomrule
\end{tabular}
\end{table}

\begin{table}[htbp]
\caption{Comparison on noisy target audios for models trained on clean (-C) and noisy (-N) training data.}
\label{tab:noisy_data}
\centering
\begin{tabular}{lccc}
\toprule
Model & Naturalness & Similarity & CER \\
\midrule
GT & 4.79$\pm$0.02 & 3.64$\pm$0.08 & 3.86$\pm$0.20 \\
Whisper-C & 3.04$\pm$0.05 & 2.99$\pm$0.08 & 7.29$\pm$0.38 \\
Ours-C & \textbf{3.57}$\pm$\textbf{0.05} & \textbf{3.16}$\pm$\textbf{0.08} & \textbf{4.56$\pm$0.25} \\
Whisper-N & 1.29$\pm$0.04 & 1.86$\pm$0.07 & 24.05$\pm$0.97 \\
Ours-N & \textbf{3.52}$\pm$\textbf{0.05} & \textbf{3.32}$\pm$\textbf{0.08} & \textbf{5.04$\pm$0.28} \\
\bottomrule
\end{tabular}
\end{table}

Results from Table \ref{tab:clean_data} and \ref{tab:noisy_data} reveal the superiority of our training approach over the baseline with Whisper transcripts not only for training from partly noisy data (compare {\bf Ours-N} to {\bf Whisper-N}), but also for training from completely clean data (original LibriTTS, VCTK and LibriLight recordings, rows {\bf Ours-C} and {\bf Whisper-C}). This could be explained by the fact that even though modern speech recognition systems reach rather high accuracies, they are still prone to errors in transcripts that may hinder the quality of TTS training. The biggest contrast between the Whisper-based and the HuBERT-based models can be seen at inference from noisy reference audios (Table \ref{tab:noisy_data}). This behavior can be attributed to lack of noisy conditioning in Whisper-based input. During training, which involves reconstruction of noisy data, the model attempts to infer noise from reference embeddings, leading to decreased noise robustness of speaker encoder.


\subsubsection{HuBERT features noise encoding ability}
To verify the ability of the HuBERT model to encode noise information, we trained a binary CatBoost \cite{catboost2018} classifier to distinguish between HuBERT features generated from non-noisy and noisy speech from a subset of ChiME3 data. We conducted this using a leave-one-out scheme on the four types of noises present in ChiME3. The resulting F-scores surpassed 0.97, indicating significant separability of HuBERT features between noisy and non-noisy speech.

\section{Conclusion}

We developed a multi-task training strategy for the speaker encoder of our zero-shot TTS system. This strategy combines a speech synthesis objective, enhancing the speaker embedding's capability to encode various speech characteristics beneficial for voice cloning, with a self-supervised DINO objective that improves the speaker representation encoding and noise robustness of our system. The conducted experiments reveal significant enhancements in similarity in both quiet and especially real-life noisy environments, compared to baselines. We also empirically verified improved style encoding through an emotion classification experiment.

Furthermore, we experimentally demonstrated the benefits of a two-stage training strategy that exploits the  capabilities of self-supervised learning content encoder (S2U) to differentiate between noisy and clean speech. This strategy comprises pretraining a text-to-HuBERT-unit model and separately training a HuBERT-unit-to-speech model. We empirically verified HuBERT's ability to distinguish between noisy and clean inputs, which enables our system to associate noisy inputs with corresponding noisy outputs during training, ensuring synthesis of clear speech from clean units during inference.

Regarding the limitations of our work, we focused primarily on traditional speaker-encoder-based TTS systems, and did not perform comparisons with state-of-the-art continuous flow-based models, such as P-Flow \cite{pflow} and VoiceBox \cite{voicebox}. Additionally, we used mBART large for the text-to-unit component of our system, which may not be the most resource-efficient option. In the future, we plan to explore our approach using smaller, more efficient text-to-unit models and consider adapting our modifications for integration with latest TTS models.

\bibliographystyle{IEEEtran}
\bibliography{mybib}

\end{document}